\documentstyle[12pt]{article}
\def\beq{\begin{equation}}
\def\eeq{\end{equation}}
\def\bea{\begin{eqnarray}}
\def\eea{\end{eqnarray}}
\begin {document}
\begin{titlepage}
December 1992 \hfill HU Berlin--IEP--92/7  \\
\mbox{ }  \hfill hep-th@xxx/9212004
\vspace{6ex}
\Large
\begin {center}
\bf{Remarks on the continuum formulation of noncritical strings}\footnote{to
appear in the proceedings of the \em XXVI. Int. Symp. on the Theory of
Elementary Particles \\
\hspace*{6cm} Wendisch-Rietz, September 1992 \em}
\end {center}
\large
\vspace{3ex}
\begin{center}
H. Dorn and H.-J. Otto \footnote{e-mail: dorn@ifh.de or otto@ifh.de}
\end{center}
\normalsize
\it
\vspace{3ex}
\begin{center}
Fachbereich Physik der Humboldt--Universit\"at \\
Institut f\"ur Elementarteilchenphysik \\
Invalidenstra\ss e 110, D--O--1040 Berlin, Germany
\end{center}
\vspace{6 ex }
\rm
\begin{center}
\bf{Abstract}
\end{center}
\vspace{3ex}
We discuss various aspects of the calculation of correlation
functions in conformal theories coupled to quantized 2-dimensional gravity.
The main emphasis lies on the construction of a continuation in the number
of insertions of the cosmological constant operator for $arbitrary$ dimension
of the target space. Following closely our paper \cite{DO2} we add a more
extended introduction and discussion of the peculiarities in
1-dimensional target space as well as some further remarks about the
4-point function.

\end {titlepage}
\newpage
\setcounter{page}{1}
\pagestyle{plain}
\section {Introduction}
We will discuss here some properties of the $N$-point correlation
function
\beq
G_{N} = \langle V_{(1)}(z_{1}).....V_{(N)}(z_{N}) \rangle
\label{1}
\eeq
in the continuum formulation based on the functional integral approach.
The $V_{(i)}$ represent primary operators in a conformal theory which
have been "dressed" after coupling to quantized 2-dimensional gravity.
The $G_{N}$'s are needed also in string theory, after integration over
the $z_{i}$'s one gets string S-matrixelements
\beq
A_{N}~=~\frac{1}{V_{CKV}}~\int \prod _{i}dz_{i}~G_{N}(z_{1},...,z_{N})~~.
\label{neu}
\eeq
The simplest case is
\beq
V_{(j)}(z) = V_{k_{j}}(z) = e^{ik_{j}X(z)}
\label{2}
\eeq
where $X(z)$ is the string position in a $d$-dimensional target space
($V_{k}(z)$ tachyon vertexoperator) or an 1-dimensional free field with
background charge (Coulomb gas description of minimal conformal models).
For \em critical \em strings the Weyl degree of freedom $e^{\phi (z)}$
( $\phi$ Liouville field ) in the 2-dimensional metrics
\beq
g_{ab} = e^{\phi (z)}\hat{g}_{ab}(z)
\label{3}
\eeq
has to decouple. This yields the well known restriction $d = 26$. On the
other side for \em noncritical \em strings $d$ is a priori arbitrary.
$\phi (z)$ couples as a quantum field and the total action is \cite{D,DK,DO1}
\begin{eqnarray}
S&=&\frac{1}{8\pi}\int d^{2}z\sqrt{\hat{g}(z)}~ \hat{g}^{mn}(z)\partial_{m}
X^{\mu}(z)\partial_{n}X_{\mu}(z) \nonumber\\
&+&\frac{1}{8\pi}\int d^{2}z\sqrt{\hat{g}(z)}\Big(\hat{g}^{mn}(z)\partial_{m}
\phi(z)\partial_{n}\phi (z)+Q \hat{R}(z)\phi(z)+\mu^{2}e^{\alpha\phi(z)}\Big)
\nonumber\\
&+& ghost~ part.
\label{4}
\end{eqnarray}
with
\begin{equation}
Q^{2}=\frac{25-d}{3}~~,~~\alpha_{\pm}=\frac{Q}{2}\pm \frac{\sqrt{Q^{2}-8}}{2},
{}~~~  \alpha\equiv \alpha_{-}~~.
\label{5}
\end{equation}
This choice of parameters ensures the vanishing of the total central charge
\beq
c_{M}~+~c_{L}~-~26~=~c_{tot}~=~0
\label{6}
\eeq
and conformal dimension $1$ for $e^{\alpha \phi}$. Then the partition function
is invariant with respect to scalings of the background metrics $\hat{g}$.
The action (\ref{4}) can be compared with that for a $(d+1)$-dimensional
string living in a general target space \cite{Ts}
\bea
S~=~\frac{1}{4\pi \alpha ^{\prime}}\int d^{2}z \sqrt{g} &\Big( &G_{MN}
(X(z))g^{mn} \partial _{m} X^{M}(z) \partial _{n} X^{N}(z)~-~\frac{i
\epsilon^{mn}}{\sqrt{g}} B_{MN} \partial _{m} X^{M} \partial _{n}
X^{N}\nonumber\\
&+& \alpha ^{\prime} R^{(2)} \Phi (X(z))~+~\alpha ^{\prime} T(X(z))\Big)~~.
\label{7}
\eea
Weyl invariance requires the renormalization group $\bar{\beta}$-functions
to vanish. Identifying
$\phi$ with the $(d+1)$ th coordinate
\beq
X^{M}~=~(~X^{\mu},\phi ~)~~,
\label{8}
\eeq
we see that on the level of the action a noncritical string in flat
$d$-dimensional space is equivalent to a (critical) string in
$(d+1)$-dimensional target space with trivial $G, B$ but nontrivial $\Phi ,
T$
\bea
\Phi (X^{M}(z))&=& \frac{1}{2} Q \phi (z)~~,\\
\label{9}
T (X^{M}(z))&=& \frac{1}{2} \mu ^{2} e^{\alpha \phi (z)}~~.
\label{10}
\eea
This describes a linear dilaton and exponential tachyon background.
While this equivalence for the action is trivial it remains an open question
for string S-matrixelements. The problem is connected with the question
whether there are indeed asymptotic states in the target space, too \cite{Hen}.

Concluding this introduction we summarize some well known facts which
one should have in mind in the discussion of the following sections.
{}From (\ref{5}) one finds $\alpha $ to be real for $d\leq 1$, complex for $1<
d<25$ and pure imaginary for $d\geq 25$. Therefore, the model at least in the
present formulation is inconsistent for $1<d<25$. For $d\geq 25$, after a
redefinition $\phi \rightarrow i \phi $, the Liouville field can be
interpreted as a time-like coordinate \cite{Inder,Pol}. The change from
$\mu ^{2}=0$ to $\mu ^{2} \neq 0$ is for all $d$ highly nontrivial. There is
no smooth limit $\mu ^{2} \rightarrow 0$, scalings in $\mu ^{2}$ can always be
compensated by a shift in $\phi $, for $\mu ^{2} \neq 0$ in contrast
to $\mu ^{2} = 0$ there is no SL(2,C) invariant vacuum \cite{Sei}.
\section{Peculiarities for 1-dimensional target space}
Although $d=1$ is unphysical from the string point of view, this case has
attracted particular attention. Besides its relevance for minimal conformal
models,
this is due to the achievements in the study of matrix models \cite{GM,DS}.
In these models one can sum the contributions of world surfaces of all
topological genera via the so called double scaling limit. To check whether
the emerging theory has indeed something to do with what one would expect to be
the continuum formulation of a conformal theory coupled to 2-dim gravity,
correlation functions have been calculated via the continuum functional
integral representation, both for $\mu ^{2}=0$ as well as for $\mu ^{2}
\neq 0$ \cite{Polya,GL,FK,Dots,ADH}.

To motivate our discussion in the following sections we now want to emphasize
the kinematical peculiarities for $d=1$. The barrier at $d=1$ mentioned
at the end of section 1 is in fact a barrier in the central charge $c_{M}$
of the matter field $X$. Adding a background charge $P^{\mu}$ for the matter
field in eq.(\ref{4}), i.e.
\beq
S~\rightarrow ~S~+~ \frac{i}{8\pi} \int d^{2}z\sqrt{\hat{g}}~ \hat{R}(z)
P^{\mu}X_{\mu}(z)~~,
\label{11}
\eeq
leads to
\beq
c_{M}~=~d-3P^{2}~~~,~~Q^{2}~=~\frac{25-c_{M}}{3}~=~\frac{25-d}{3}+P^{2}
\label{12}
\eeq
with unchanged $\alpha (Q)$ relation,(\ref{5}). The gravitational dressed
tachyon vertex operator \cite{D,DK,DO1} is
\beq
V_{k} = e^{ikX(z)}e^{\beta\phi(z)}
\label{13}
\eeq
with
\beq
\beta = \frac{Q}{2} - \sqrt{m^{2}+(k-P/2)^{2}}
\label{14}
\eeq
\beq
m^{2} \equiv \frac{1-d}{12}~~.
\label{15}
\eeq
$\beta$ is real for most of the $k$, but there is an exceptional compact
domain $E$ whose boundary is given by $\beta =\frac{Q}{2}$ and which is
characterized by
\bea
m^{2}+k_{\perp}^{2}&<&0,
\nonumber\\
-\sqrt{-m^{2}-k_{\perp}^{2}}&<&k_{\parallel}-\frac{\vert P \vert }{2}<
+\sqrt{-m^{2}-k_{\perp}^{2}}
\label{16}
\eea
with $k_{\parallel},k_{\perp}$ denoting the components of $d$-dimensional
momentum $k$ parallel and orthogonal to $P$. We start out with $k \notin E$
and continue afterwards.

Let us consider for the moment the simpler case $\mu ^{2} =0$. Then the
integrations in the functional integral for $A_{N}$ are all of Gaussian type
and we find e.g. for $A_{4}$ at arbitrary $d$ up to trivial numerical factors
\beq
A_{4}~=~\frac{\Gamma (\frac{S+m^{2}}{2})}{\Gamma (1-\frac{S+m^{2}}{2})}
\frac{\Gamma (\frac{T+m^{2}}{2})}{\Gamma (1-\frac{T+m^{2}}{2})}
\frac{\Gamma (\frac{U+m^{2}}{2})}{\Gamma (1-\frac{U+m^{2}}{2})}~~.
\label{17}
\eeq
This is just the $(d+1)$ -dimensional critical Virasoro-Shapiro amplitude
with a slightly modified definition of the Mandelstam variables
\bea
S~=~\Big (k_{1}+k_{2}-\frac{P}{2}\Big )^{2}-\Big (\beta _{1}+\beta_{2}-
\frac{Q}{2}\Big )^{2}
\nonumber\\
T~=~\Big (k_{1}+k_{3}-\frac{P}{2}\Big )^{2}-\Big (\beta _{1}+\beta_{3}-
\frac{Q}{2}\Big )^{2}
\nonumber\\
U~=~\Big (k_{1}+k_{4}-\frac{P}{2}\Big )^{2}-\Big (\beta _{1}+\beta_{4}-
\frac{Q}{2}\Big )^{2}
\label{18}
\eea
and
\beq
\sum _{i}k_{i}=P~~,~\sum _{i}\beta _{i}=Q~.
\label{19}
\eeq
Now for $d=1$ i.e. $m^{2}=0$ the $\beta (k)$ relation linearizes
\beq
\beta ~=~\frac{Q}{2}~-~\epsilon _{i}\Big (k_{i}-\frac{P}{2}\Big )~~,~~
\epsilon _{i}~=~sgn\Big (k_{i}-\frac{P}{2}\Big )~~.
\label{20}
\eeq
The sum rules for the $k$'s and $\beta $'s (\ref{19}) imply separate
sum rules for the $k_{i}$ with $\epsilon _{i}=+1$ and the $k_{i}$
with $\epsilon _{i}=-1$ \cite{FK} . Not all $\epsilon _{i}$ can be
equal. In the case $(-+++)$ one finds e.g.
\beq
k_{1}~=~-\frac{Q}{2}~~,~~k_{2}+k_{3}+k_{4}~=~\frac{Q}{2}+P~~.
\label{21}
\eeq
This leads to
\bea
S~=~\frac{(Q+P)^{2}}{4}~-~k_{2}(Q+P) \nonumber \\
T~=~\frac{(Q+P)^{2}}{4}~-~k_{3}(Q+P) \nonumber \\
U~=~\frac{(Q+P)^{2}}{4}~-~k_{4}(Q+P)
\label{22}
\eea
and therefore a factorization of the amplitude into "leg-factors" depending
on a single momentum only. To understand the infinite number of poles in
spite of the absence of transversal degrees of
freedom at $(1+1)$ dimensions one has to perform a complete BRST analysis
\cite{W,Bouw}.
\section{Three point function for $d>1~,~\mu ^{2}\neq 0$}
The calculation of correlation functions for the nontrivial case of
nonvanishing cosmological constant $\mu^{2}$ has been done by continuation
in the quantity $s$ defined below. In this process extensive use has been made
of the relations among momenta due to the kinematical peculiarities in $d=1$
dimensional target space \cite{GL,FK,Dots,ADH}. We now want to get rid of
this constraint and follow closely our paper \cite{DO2}.

The three point function has been calculated in \cite{GL,FK} for positive
integer $s$
\beq
\alpha s =Q-\sum_{j=1}^{3}\beta_{j}
\label{e8}
\eeq
\bea
A_{3}=\delta (\sum_{j}k_{j}-P)\frac{\Gamma (-s)}{\alpha}\Gamma (1+s)
\Big(\frac{\mu ^{2}}{8}\frac{\Gamma (1+\frac{\alpha ^{2}}{2})}{
\Gamma (-\frac{\alpha ^{2}}{2})}
\Big)^{s}~~\times ~~~~~~~~~~~~~~~~~~~~~~~~~~~~~~~~~~~~~~~\nonumber\\
exp\Big \{f(-\frac{\alpha ^{2}}{2},-\frac{\alpha ^{2}}{2}\vert s)-
f(1+\frac{\alpha ^{2}}{2},\frac{\alpha ^{2}}{2}\vert s)
+\sum_{j=1}^{3}\Big(
f(1-\alpha \beta_{j},-\frac{\alpha ^{2}}{2}\vert s)-
f(\alpha \beta_{j},\frac{\alpha ^{2}}{2}\vert s)\Big)\Big\}.
\label{e9}
\eea
The infinity due to the pole in $\Gamma(-s)$ is understood as a signal of the
logarithmic modification of the $\mu $ -scaling law for positive integer $s$
\cite{FK,DO1}. We introduced $f(a,b\vert s)$ as
\beq
f(a,b\vert s)=\sum_{j=0}^{s-1}\log \Gamma (a+bj).
\label{e10}
\eeq
This function fulfills the relations
\beq
f(a,b\vert s+1)=f(a,b\vert s)+\log \Gamma (a+bs)~~,
\label{e11}
\eeq
\beq
f(a+1,b\vert s)=f(a,b\vert s)+s\log b +\log \Gamma (\frac{a}{b}+s)-
\log \Gamma (\frac{a}{b})~~,
\label{e12}
\eeq
\beq
f(a+b,b\vert s)=f(a,b\vert s)+\log \Gamma (a+bs)-\log \Gamma (a)~~,
\label{e13}
\eeq
\beq
f(a+b(s-1),-b\vert s)=f(a,b\vert s)~~,
\label{e14}
\eeq
\beq
f(2a,2b\vert s)=f(a,b\vert s)+f(a+\frac{1}{2},b\vert s)+
(s(s-1)b+s(2a-1))\log 2 -\frac{s}{2}\log \pi~~,
\label{e15}
\eeq
\beq
f(a,0\vert s)=s\log \Gamma (a)~~,
\label{e16}
\eeq
\beq
f(a,\frac{1}{s}\vert s)=\frac{1}{2}(s-1)\log 2\pi+(\frac{1}{2}-sa)\log s
+\log \Gamma(sa).
\label{e17}
\eeq
We have found a continuation of $f$ to arbitrary $s$, which is given by the
following integral representation
\bea
f(a,b\vert s)=\int_{0}^{\infty }\frac{dt}{t}\Big(s(a-1)
e^{-t}+b\frac{s(s-1)}{2}
e^{-t}-s\frac{e^{-t}}{1-e^{-t}}
\nonumber\\
+\frac{(1-e^{-tbs})e^{-at}}{(1-e^{-tb})
(1-e^{-t})}\Big).
\label{e18}
\eea
By careful manipulations with this integral representation one can show, that
our continuation fulfills all the functional relations (\ref{e11}-\ref{e17})
also for generic complex $a,b,s$.
The integral representation (\ref{e18}) is convergent for positive real parts
of $a,b,s$. Together with the functional relations this fixes the complete
analytical structure of $f(a,b\vert s)$. Due to (\ref{e14}) we can restrict
ourselves to $Re b>0$ . Then we find via (\ref{e11}) and (\ref{e12}) that
$exp(f(a,b\vert s)$ has poles at
\beq
a=-bj-l~~~~~~~(poles)
\label{e19}
\eeq
and zeros at
\beq
a+bs=-bj-l~~~~(zeros)~~,
\label{e20}
\eeq
with $Re ~b>0;~~j,l=0,1,2,...~.$
The order of poles and zeros is determined by the number of different
realizations of the r.h.s. of eqs. (\ref{e19}) and (\ref{e20}),
respectively.
Using (\ref{e14}) we get from (\ref{e9})
\beq
A_{3}=\delta (\sum_{j}k_{j}-P)\frac{\Gamma (-s)}{\alpha}\Gamma (1+s)
\Big(\frac{\mu ^{2}}{8}\frac{\Gamma (1+\frac{\alpha ^{2}}{2})}{
\Gamma (-\frac{\alpha ^{2}}{2})}
\Big)^{s}
\prod_{i=0}^{3}F_{i}~~,
\label{e21}
\eeq
where we introduced
\bea
\overline {\beta_{i}}\equiv 1/2(\beta _{j}+\beta _{k}-\beta _{i})=
1/2(Q-\alpha s)-\beta _{i}~~,
\nonumber\\
(i,j,k)~=~perm(1,2,3)~~,
\label{meyer1}
\eea
\beq
F_{i}~=~exp\{ f(\alpha \overline{\beta _{i}},\frac{\alpha ^{2}}{2}\vert s)
-f(\alpha \beta _{i},\frac{\alpha ^{2}}{2}\vert s)\}
\label{e22}
\eeq
and
\beq
\alpha \beta _{0}~=~1+\frac{\alpha ^{2}}{2},~~~~~\alpha \overline{\beta _{0}
}~=~-\frac{\alpha ^{2}s}{2}.
\label{e23}
\eeq
Eqs. (\ref{e19}) and (\ref{e20}) determine the poles and zeros of $F_{i}$
as follows
\beq
2\beta _{i}~=~\alpha _{+}l_{i}~+~\alpha _{-}(j_{i}-s)~~~(poles),
\label{e24}
\eeq
\beq
2\beta _{i}~=~\alpha _{+}l_{i}~+~\alpha _{-}j_{i}~~~~~~~(zeros).
\label{e25}
\eeq
The integers $j_{i},l_{i}$ have to be both $\leq 0$ or both $>0$, i.e.
\beq
(j_{i}~-~1/2)~(l_{i}~-~1/2)~>~0.
\label{e26}
\eeq
For general $s$ poles and zeros are located at different values of
$\beta _{i}$, but if $\alpha _{-}s~=~\alpha _{-}m~+~\alpha _{+}n$ with integer
$m,n$
a lot of poles and zeros cancel.

Of special interest for our later discussion is integer $s>0$. In this this
case the remaining pole-zero pattern is given by
\beq
2\beta _{i}~=~\alpha _{+}(l_{i}+1)~-~\alpha _{-}j_{i}~~~(poles)~~,
\label{e27}
\eeq
\beq
2\beta _{i}~=~-\alpha _{+}l_{i}~-~\alpha _{-}j_{i}~~~~~~(zeros)~~,
\label{e28}
\eeq
with
\beq
integer~s~\geq 0;~~~~j_{i}~=~0,1,...,s-1;~~~~l_{i}~=~0,1,...\infty.
\label{e29}
\eeq
In contrast to the general case the pole positions in $\beta _{i}$
are bounded from below.

Such boundedness is achieved also for $s>0$ with
\beq
\alpha _{-}s~=~\alpha _{+}n~+~\alpha _{-}([s]-m);~~~~1~\leq ~n~\leq ~m~
\leq ~[s].
\label{meyer2}
\eeq

Special care is needed for $F_{0}$. Due to (\ref{e23}) for all $s$ we
are just sitting on the $j_{0}=l_{0}=1$ zero of (\ref{e25}). On the other
hand, for positive integer $s$, where we start our continuation, due to
pole-zero cancellation $F_{0}$ is finite. A finite $F_{0}$ emerges also in
the more general situation
\beq
\alpha _{-}s~=~\alpha _{+}(l_{0}-1)~+~\alpha _{-}(j_{0}-1)~~,
\label{meyer3}
\eeq
with $l_{0},j_{0}$ fulfilling (\ref{e26}).

In all cases where pole-zero cancellation does not guarantee finite $F_{0}$
we treat it using (\ref{e12}) as
\beq
F_{0}~=~-(\frac{\alpha ^{2}}{2})^{-s}\frac{\pi}{\Gamma (1+s)}exp~\Big[ f(1
-\frac{\alpha ^{2}}{2}s,\frac{\alpha ^{2}}{2}\vert s)-f(1+\frac{\alpha
^{2}}{2},
\frac{\alpha ^{2}}{2}\vert s)\Big] ,
\label{e30}
\eeq
where a factor $\Gamma (0) sin~\pi s$ has been set equal to 1.
Of course, this procedure still has to be justified by comparing the $s$-
asymptotics of the resulting expression with that of the original functional
integral. (For $d=1$ see ref \cite{ADH}.)

The analytic structure af $A_{3}$ beyond the trivial $s$-dependence in $F_{0}$
and the first factors in (\ref{e21}) is determined by that of $\prod_{i=1}
^{3}F_{i}$. Eliminating $s$ in (\ref{e24}) by using (\ref{meyer1}) we get
finally
\beq
2\overline{\beta _{i}}=\alpha _{-}j_{i}+\alpha _{+}l_{i}~~~~~~~~~~~~(poles)~~,
\label{e31}
\eeq
\beq
2\beta _{i}=\alpha _{-}j_{i}+\alpha _{+}l_{i}~~~~~~~~~~~~(zeros)~~,
\label{e32}
\eeq
with $j_{i},l_{i}$ respecting (\ref{e26}).

In contrast to the $d=1$ case, which we are going to discuss below,
for general $d>1$ the pole structure does not factorize into leg
poles. An attempt to interprete the pole structure will be made in section 4.

For $d>1$ the three $\beta _{i}$'s are independent. For $d=1$ there remains no
angular degree of freedom in $k$-space. The defining equation (\ref{14})
reduces to eq. (\ref{20}).
Then momentum conservation $\sum k_{i}=P$ induces constraints among the
$\beta _{i}$'s \cite{FK}. \footnote{Remember that for $\mu ^{2} \neq 0$ there
is
$no$ sum rule $\sum \beta _{i}=Q$.}
The $\epsilon _{i}$ must be either all equal to one another but opposite
to $sign(P)$ or of the type $(++-)$ or $(--+)$: One finds in the first case
for $i$=1,2,3
\beq
2\overline{\beta _{i}}=-2\beta _{i}+\frac{3Q+P}{2}~~.
\label{e34}
\eeq
For $(++-)$ one gets instead
\beq
2\overline{\beta _{1}}=2\beta _{2}-\frac{Q+P}{2},~~~
2\overline{\beta _{2}}=2\beta _{1}-\frac{Q+P}{2},~~~
2\overline{\beta _{3}}=\frac{Q+P}{2}~~.
\label{e35}
\eeq
Finally, for the case $(--+)$ one has to replace in (\ref{e35}) $P$ by $-P$.

In all cases the position of any pole or zero of $\prod_{i=1}^{3}F_{i}$
becomes a function of a single $\beta_{i}$. To be more explicit, we
consider the situation studied in detail in \cite{FK}, i.e. $(++-)$ and $P<0$.
Then $2\alpha =Q+P$ and using eqs.(\ref{e12}),(\ref{e13}),(\ref{e22})
and (\ref{e30}) we get from
(\ref{e21})
\beq
A_{3}=-\pi \delta (\sum k_{i}-P)\frac{1}{\alpha }
\Big(-\frac{\alpha ^{2}\mu ^{2}}{16}\Delta (\frac{\alpha ^{2}}{2})\Big)^{s}
\prod_{i=1}^{3}\Delta(m_{(i)})~~,
\label{e36}
\eeq
with $\Delta (x)=\frac{\Gamma (x)}{\Gamma (1-x)},~~
m_{(i)}=(\beta _{i}^{2}-k_{i}^{2})/2$. This coincides with \cite{FK} up to a
trivial factor $-\pi (-\frac{\alpha ^{2}}{2})^{s}$.

\section{Four point function for $d>1~,~\mu ^{2} \neq 0$}
Performing the functional integral using the technique of refs. \cite{GL,FK}
one finds for integer $s_{4} \geq 0$
\beq
\alpha s_{4}~=~ Q-\sum _{i=1}^{4} \beta _{i}
\label{gl1}
\eeq
\bea
A_{4}~=&~&\delta (\sum k_{j}-P) \frac{\Gamma (-s_{4})}{\alpha} \Big (\frac
{\mu ^{2}}{8\pi}\Big )^{s_{4}} \nonumber \\
&~& \int  d^{2}z \prod _{I=1}^{s_{4}}d^{2}w_{I} \vert z \vert ^{2k_{4}k_{1}-
2\beta _{4} \beta _{1}} \vert 1-z \vert ^{2k_{4}k_{2}-2\beta _{4} \beta _{2}}
\prod _{1\leq I<J\leq s_{4}} \vert w_{I}-w_{J} \vert^{-2\alpha ^{2}} \nonumber
\\&~& \prod _{I=1}^{s_{4}} \vert z-w_{I} \vert ^{-2\alpha \beta _{4}}
\vert w_{I} \vert ^{-2\alpha \beta _{1}} \vert 1-w_{I} \vert ^{-2 \alpha \beta
_{2}}~~.
\label{gl2}
\eea
Unfortunately up to now there is no generalization of the Dotsenko-Fateev
integral formula \cite{DF} to more general exponents as needed in (\ref{gl2}).
The determination of $A_{4}$ via functional relations in \cite{FK} makes
extensive use of the relations between the exponents induced by the peculiar
kinematics in $d=1$ dimensions.

Therefore, as a first step we consider special 4-point functions which,
hopefully, still allow an exploration of the general singularity structure
in a definite Mandelstam channel. The first possibility is
\beq
A_{4}(k_{1},k_{2},k_{3},k_{4}=0)~=~\frac{d}{d\mu ^{2}}A_{3}(k_{1},k_{2},
k_{3})~=~\frac{s}{\mu ^{2}}A_{3}(k_{1},k_{2},k_{3})~~,
\label{gl3}
\eeq
the second simple possibility is based on the requirement $\beta _{4}=0$.
This decouples the $z$ and $w$ integrals resulting in ($\tilde{A}_{3}$
is the 3-point function $A_{3}$ without the momentum conservation
$\delta$ -function)
\bea
A_{4}(k_{1},k_{2},k_{3},k_{4})~=&&\delta(\sum _{i=1}^{4}k_{i}-P)~\tilde{A}_{3}
(k_{1},k_{2},k_{3})\nonumber\\
&& \pi ~\frac{\Gamma(1+k_{4}k_{1})}{\Gamma (-k_{4}k_{1})}~ \frac{\Gamma (1+
k_{4}k_{2})}{\Gamma (-k_{4}k_{2})} ~\frac{\Gamma (-1-k_{4}k_{1}-k_{4}k_{2})}
{\Gamma (2+k_{4}k_{1}+k_{4}k_{2})}
\label{gl4}
\eea
with
\beq
(k_{4}-\frac{P}{2})^{2}~=~\frac{Q^{2}}{4}-m^{2}~~.
\label{gl5}
\eeq
Using (\ref{14}) and $\sum k_{i}=P$ one finds for arbitrary $k_{i}$ the
relation
\beq
2(k_{2}k_{3}+k_{2}k_{4}+k_{3}k_{4})~=~(\beta _{1}-\frac{Q}{2})^{2}-
\sum _{i=2,3,4}(\beta _{i}-\frac{Q}{2})^{2}+2m^{2}+\frac{P^{2}}{2}
\label{gl6}
\eeq
and 3 further eqs. for cyclic permutations of the indices.

In the case $k_{4}=0$ this means ($(i,j,l)=perm(1,2,3)$)
\beq
2k_{i}k_{j}=
(\beta _{l}-\frac{Q}{2})^{2}-(\beta _{i}-\frac{Q}{2})^{2}-
(\beta _{j}-\frac{Q}{2})^{2}+m^{2}+\frac{P^{2}}{4}~~.
\label{gl7}
\eeq
In the other case $\beta _{4}=0$ as in general the products $k_{i}k_{j}$
are not fixed by the $\beta _{i}$.

We now turn to the discussion of $k_{4}=0$ i.e. expression (\ref{gl3}).
The relation (\ref{gl7}) then implies that any candidate for Mandelstam
variables in this special situation can be expressed in terms of $\beta _{1},
\beta _{2}, \beta _{3}$.
If one defines with a $d$-dimensional picture in mind
\beq
t_{12}=(k_{1}+k_{2}-\frac{P}{2})^{2}=(k_{3}-\frac{P}{2})^{2}=t_{34}
\label{e39}
\eeq
etc., the pole positions (\ref{e31}) are determined by
\beq
\sqrt{t_{23}+m^{2}}-\sqrt{t_{13}+m^{2}}-\sqrt{t_{12}+m^{2}}=
\alpha _{-}(j-\frac{1}{2})+\alpha _{+}(l-\frac{1}{2})~~.
\label{e40}
\eeq
To get a more familiar picture we now use the correspondence
to $(d+1)$-dimensional critical strings \cite{FK}. For $\mu^{2} =0$
the amplitudes are equal to that of a $(d+1)$-dimensional critical string
theory with a conservation law
$$
\sum k_{i}=P,~~~~\sum \beta_{i}=Q
$$
and eq. (\ref{14}) can be interpreted as a mass shell condition
\beq
(\beta -\frac{Q}{2})^{2}-(k-\frac{P}{2})^{2}=m^{2}~~.
\label{e41}
\eeq
The corresponding Mandelstam variables are
$$
T_{ij}^{0}=(\beta _{i}+\beta _{j}-\frac{Q}{2})^{2}-
(k_{i}+k_{j}-\frac{P}{2})^{2}~~.
$$
Also for $\mu ^{2} \neq 0$ but integer $s_{4}\geq 0$, with
\beq
\alpha s_{4}=Q-\sum_{i=1}^{4}\beta _{i}~~,
\label{e42}
\eeq
a $(d+1)$-dimensional interpretation is possible. The 4-point function then
turns into a $(4+s_{4})$-point function with $k_{5}=k_{6}=...=k_{4+s_{4}}=0$.
Motivated by these special cases
we define in general
\beq
T_{ij}=(\beta _{i}+\beta _{j}-\frac{Q-\alpha s_{4}}{2})^{2}-
(k_{i}+k_{j}-\frac{P}{2})^{2}.
\label{e43}
\eeq
This ensures $T_{12}=T_{34}$ etc. and fulfills
\beq
T_{12}+T_{13}+T_{14}=4m^{2}-\frac{1}{4}(Q^{2}-P^{2}+\alpha s_{4}
(\alpha s_{4}+2Q)).
\label{e44}
\eeq
Applying this general definition to our special case (\ref{gl3}) we get
$(k_{4}=0, \beta _{4}=\alpha, s_{4}=s-1)$
\beq
T_{i4}~=~m^{2}-\alpha (1+\frac{s_{4}}{2})(2\overline{\beta _{i}}+
\frac{\alpha s_{4}}{2})~;~~~~~i=1,2,3.
\label{e45}
\eeq
This gives a transcription of the pole structure (\ref{e31}) of the
3-point function into poles in the Mandelstam variables (\ref{e43})
of the (special) 4-point function. At least under the constraint of
fixed $ T_{14}+T_{24}+T_{34} $, i.e. fixed $ s_{4}=s-1 $ the poles (\ref
{e31}) turn directly into poles in the individual Mandelstam variables.
The resulting spectrum is for general $s$ unbounded in both directions.

For the particular case of fixed integer $s\geq 0$, due to the pole-zero
cancellation leading to (\ref{e27}),(\ref{e28}) we find poles in e.g.
$T_{12}=T_{34}$ only for
\beq
T_{34}~=~m^{2}~+~(2+s_{4})\Big(l-\frac{\alpha ^{2}}{2}(j-\frac{s_{4}}{2})\Big).
\label{e46}
\eeq
These poles can be seen to be due to the standard poles at $m^{2}+2l,
{}~integer~ l\geq 0$ in those channels of the $(4+s_{4})$-point amplitude
which are
constituted by the momenta $k_{1}, k_{2}$ and $j$ times $k=0$.
The pole spectrum in $T_{34}$ is bounded from below also for values of
$s$ fulfilling (\ref{meyer2}). It is remarkable that than in addition
$F_{0}$ is well defined without any modification.

For the other special case $\beta _{4}=0$  a $(d+1)$-dimensional point of
view seems to be doubtful from the very beginning. The factor $\tilde{A}_{3}$
depends on the momenta only via $\beta _{i} (i=1,2,3)$, but now the $\beta
^{\prime}$s cannot be expressed in terms of the Mandelstam variables. In
contrast to the
case before we are not in the rest frame of one particle.

Defining as in (\ref{e39}) $d$-dimensional Mandelstam variables $t_{ij}$
we find
\bea
A_{4}~=~\delta \Big (\sum _{i=1}^{4}k_{i}-P\Big )~\pi
{}~\tilde{A}_{3}(k_{1},k_{2}
,k_{3})~
\prod _{i=1}^{3} \frac{\Gamma (\frac{t_{4i}+m^{2}_{i}}{2})}{\Gamma (1-\frac {
t_{4i}+m^{2}_{i}}{2})}\\
\label{gl8}
with~~ \Big (k_{i}-\frac{P}{2}\Big )^{2}~=~\Big (\beta _{i}-\frac{Q}{2}\Big )
^{2}-m^{2}~=~-m_{i}^{2}~~. \nonumber
\eea
The $\beta _{i}$ parametrize the masses. Any quantization condition for
the Liouville momenta $\beta _{i}$ generates a discrete set of masses
$m^{2}_{i}$. The pole spectrum in $t_{4i}$ is
\beq
t_{4i}~=~-m_{i}^{2}-2n~,~~~~n~~integer~\geq 0~~.
\label{gl9}
\eeq
This generates a constraint on the allowed values for the $hypothetically $
quantized values for $\beta $
\beq
\beta _{i}^{2}~-~\beta _{i} Q~-~\beta _{j}^{2}~+~\beta _{j} Q~=~even~integer~~.
\label{gl10}
\eeq
\section{Conclusions}
We have pointed out the role of the special kinematics at $d=1$ in the
continuation from integer to noninteger along the lines of ref.
\cite{GL,FK,Dots}
and presented a continuation $without$ the constraint $d=1$.
Since the continuation from a discrete set of points is mathematically
ambiguous our procedure still has to be proven to yield the correct
solution for the physical problem under discussion. This can be done
by comparing the $\vert s \vert \rightarrow \infty $ asymptotics with
that of the original functional integral. The corresponding problem
in one dimensional target space has been solved in ref.\cite{ADH}.
Unfortunately the technique of that paper cannot directly be applied
here as it depends essentially on the discreteness of the momentum
spectrum in minimal models.

Nevertheless we are confident to have presented a correct continuation
procedure. First for $d=1$ it reduces to that of refs. \cite{GL}, \cite{FK},
\cite{ADH} and second, the resulting analytical structure is already
completely determined by the functional equations for
$f$ and the absence of singularities for $Re~(a,b,s)\geq 0$.

The interpretation of the spectrum is still not completely clear. While
the pole structure for integer $s\geq0$ fits into the $(d+1)$-dimensional
critical string picture well known from the case of vanishing Liouville
mass $\mu$, for general $s$ the pole-zero cancellation enabling just
this interpretation is removed and an unacceptable spectrum emerges.
This may either signal a hidden disease of our model or only a breakdown
of its $(d+1)$-dimensional interpretation. Favouring the last alternative
one has to analyse eq. (\ref{e40}) and to think about mass shell conditions
in $d$-dimensional space induced by quantization of the $\beta _{i}$
values due to the Liouville dynamics. A possible guess would be to restrict
$\beta _{i}$ to the sum of integer multiples of $\alpha _{+}$ and $\alpha
_{-}$. Then our results concerning the 3-point function generate some
kind of fusion rules which have to be consistent with the pole pattern
found for the (special) 4-point functions (\ref{gl9}),(\ref{gl10}).
One further interesting application of the presented continuation
procedure concerns the off shell amplitudes for critical strings proposed
in ref. \cite{MP} recently.

\end{document}